\documentclass{article}
\usepackage{bm}
\usepackage{amsmath,amssymb}
\usepackage[top=20truemm,bottom=20truemm,left=20truemm,right=20truemm]{geometry}
\usepackage{authblk}
\usepackage[backend=biber,style=phys,articletitle=false,biblabel=brackets,chaptertitle=false,pageranges=false,uniquelist=false,eprint=true,maxbibnames=3]{biblatex}
\usepackage{hyperref}
\usepackage{xurl}
\usepackage{breqn} 
\hypersetup{breaklinks=true}
\DeclareSourcemap{
	\maps[datatype=bibtex, overwrite=true]{
		\map{
			\step[fieldset=doi, null]
			\step[fieldsource=journal,
				match={Physical Review D},
				replace={Phys. Rev. D}
			]
			\step[fieldsource=journal,
				match={Philosophical Transactions of the Royal Society A},
				replace={Philos. T. R. Soc. A}
			]
			\step[fieldsource=journal,
				match={The Astrophysical Journal Letters},
				replace={Astrophys. J. Lett.}
			]
			\step[fieldsource=journal,
				match={The Astrophysical Journal},
				replace={Astrophys. J.}
			]
			\step[fieldsource=journal,
				match={Classical and Quantum Gravity},
				replace={Classical Quant. Grav.}
			]
			\step[fieldsource=journal,
				match={Journal of High Energy Physics},
				replace={JHEP}
			]
			\step[fieldsource=journal,
				match={Proceedings of the American Mathematical Society},
				replace={P. Am. Math. Soc.}
			]
			\step[fieldsource=journal,
				match={Journal of Physics A: Mathematical and Theoretical},
				replace={J. Phys. A-Math. Theor.}
			]
			\step[fieldsource=journal,
				match={Journal of Physics A: Mathematical and General},
				replace={J. Phys. A-Math. Gen.}
			]
			\step[fieldsource=journal,
				match={Inventiones mathematicae},
				replace={Invent. Math.}
			]
			\step[fieldsource=journal,
				match={Journal of Geometry and Physics},
				replace={J. Geom. Phys.}
			]
			\step[fieldsource=journal,
				match={Journal of the Faculty of Science, the University of Tokyo. Sect. 1 A, Mathematics},
				replace={J. Fac. Sci. U. Tokyo 1}
			]
			\step[fieldsource=journal,
				match={Compositio Mathematica},
				replace={Compos. Math.}
			]
			\step[fieldsource=journal,
				match={Proceedings of the Japan Academy, Series A, Mathematical Sciences},
				replace={P. Jpn. Acad. A-Math.}
			]
			\step[fieldsource=journal,
				match={Journal of Computational and Applied Mathematics},
				replace={J. Comput. Appl. Math.}
			]
			\step[fieldsource=journal,
				match={Constructive Approximation},
				replace={Constr. Approx.}
		]
			\step[fieldsource=journal,
				match={The European Physical Journal C},
				replace={Eur. Phys. J. C}
			]
			\step[fieldsource=journal,
				match={General Relativity and Gravitation},
				replace={Gen. Relativ. Gravit.}
			]
		}
	}
}

\makeatletter
	
		\@addtoreset{equation}{section}
\makeatother
\title{Strong deflection limit-analysis using Picard-Fuchs equation in Einstein-Maxwell-Dilaton spacetime}
\date{\today}
\author{Tadashi Sasaki\thanks{ta-sasaki@kumagaku.ac.jp}}
\affil{Faculty of Commerce, Kumamoto Gakuen University, Kumamoto 862-8680, Japan}
\begin{document}

\maketitle

\begin{abstract}
{{{
	We consider the deflection of light by a spherically symmetric and electrically charged black hole solution
	in the Einstein-Maxwell-Dilaton theory with specific values of the dilaton coupling constant where the deflection angle
	can be represented as elliptic integrals.
	We show that the deflection angle as a function of two dimensionless variables $(s,z)$, which are related to 
	the background charge and the impact parameter, respectively, satisfies a system of 2nd-order linear 
	partial differential equations called the Picard-Fuchs (PF) equations.
	For each case of the dilaton coupling, the PF equations lead to 
	1st-order ordinary differential equations with respect to the variable $s$ for the constants $\bar{a}$ and $\bar{b}$ 
	in the log-formula for the strong deflection limit.
	Using the Hamiltonian system associated with Painlev\'{e} VI equations, 
	which holds as a result of the integrability of the PF equations,
	we solve the equations for $\bar{a}$ and $\bar{b}$.
	By requiring consistency with the Schwarzschild case in the zero-charge limit, 
	$\bar{a}$ and $\bar{b}$ for nonzero charge are uniquely determined.
}}}
\end{abstract}


\section{Introduction}
{{{
	Deflection of light due to a strong gravitational field produced by a massive compact object has been studied intensively.
	If the massive object is compact enough, there may exist topologically circular photon orbits.
	In particular, if the spacetime is spherically symmetric and the circular photon orbits are unstable in the radial direction, 
	such orbits form a spherical surface called a photon sphere.
	The existence of a photon sphere (or its generalization to axisymmetric cases) is a universal feature of stationary and asymptotically flat spacetimes representing black holes\cite{CunhaHerdeiro2020}
	and leads to the associated shadow region, 
	which has been observed at the centers of galaxies by the Event Horizon Telescope Collaboration\cite{EHTC_Akiyamaea2019,EHTC_Akiyamaea2022}.
	However, since exotic spacetimes other than black holes, such as naked singularities or wormholes, 
	can also possess a photon sphere as well as a shadow region\cite{VirbhadraEllis2002,ChibaKimura2017,Tsukamoto2020,Tsukamoto2016,Tsukamoto2021a,SoaresPereiraea2025},
	the possibility that the supermassive black hole candidates are such objects is not ruled out\cite{BambiFreeseea2019,KocherlakotaRezzollaea2021,EHTC_Akiyamaea2022a,VagnozziRoyea2023,TsukamotoKase2024,KhodadiVagnozziea2024}.

	In the geometric optics approximation, light rays, or trajectories of photons, are described by the null geodesic equations for a given background spacetime.
	Light rays passing near a photon sphere are strongly deflected and wind around the massive object an arbitrary number of times before escaping to infinity,
	which implies that the associated deflection angle $\hat{\alpha}$ diverges.
	In general, this divergence can be approximated by the following logarithmic formula\cite{Bozza2002,Tsukamoto2017}\footnote{
		The inverse relation of this formula for a Schwarzschild black hole can be found in Ref.\cite{Luminet1979}.
	}:
	\begin{equation}
		\hat{\alpha}\sim-\bar{a}\log\left(\frac{b}{b_{\rm c}}-1\right)+\bar{b},\ b\to b_{\rm c}+0 \label{SDL_Bozza}
	\end{equation}
	where $b$ is the impact parameter of the trajectory, and $b_{\rm c}$ is its critical value such that 
	the photon with the corresponding trajectory is captured by the photon sphere.
	$\bar{a}$ and $\bar{b}$ are constants determined by the background spacetime.
	We call the limit $b\to b_{\rm c}+0$ the strong deflection limit (SDL) (from outside of the photon sphere)\footnote{
		The SDL from inside of the photon sphere $b\to b_{\rm c}-0$ is possible for some exotic spacetimes\cite{ShaikhBanerjeeea2019a,Tsukamoto2021,Tsukamoto2021b},
		which we do not investigate in this paper.
	},
	and accordingly call $\bar{a}$ and $\bar{b}$ the SDL coefficients.
	The SDL coefficients appear in observable quantities such as the positions and the magnifications of the relativistic images in gravitational lensing problems\cite{VirbhadraEllis2000,Bozza2002,EiroaTorres2004}.
	Among the SDL coefficients, the logarithmic divergence rate $\bar{a}$ is totally determined by local geometrical quantities, 
	namely the areal radius of and the spacetime curvature at the photon sphere\cite{Igata2026}.
	Thus, one can immediately obtain the exact expression of $\bar{a}$ once the spacetime metric is given.
	Furthermore, $\bar{a}$ is related to the quasinormal mode frequencies of the given spacetime through the Lyapunov exponent of the unstable circular photon orbits\cite{StefanovYazadjievea2010,CardosoMirandaea2009}.
	On the other hand, the constant offset $\bar{b}$ doesn't allow such a local description.

	Usually, studies evaluating the SDL coefficients\cite{Bhadra2003,GyulchevYazadjiev,GyulchevYazadjiev2007,Tsukamoto2016,Tsukamoto2017,TsukamotoHarada2017,ShaikhBanerjeeea2019,Tsukamoto2021a,Tsukamoto2021b,JuniorJuniorea2024} have followed the method used in Ref.\cite{Bozza2002}, 
	where the integral representation of the deflection angle is divided into two parts:
	the first part, denoted by $I_{\rm D}$, diverges in the same manner as the original integral, but is given by an approximation that can be 
	evaluated using elementary functions.
	The second part, denoted by $I_{\rm R}$, is then given by the finite remainder.
	The coefficient $\bar{a}$ can be calculated solely from $I_{\rm D}$, while $\bar{b}$ has contributions both from $I_{\rm D}$ and $I_{\rm R}$.
	In general, the finite integral of $I_{\rm R}$ in the SDL is difficult to perform analytically to determine an explicit expression of $\bar{b}$
	in terms of the background parameters.
	Note that this difficulty in evaluating $\bar{b}$ is, at least partially, a technical problem.
	For example, Ref.\cite{TsukamotoGong2017,Tsukamoto2017} succeeded in obtaining an exact and analytic expression of $\bar{b}$ 
	in a Reissner-Nordstr\"{o}m spacetime by using a different integration variable from the one used before.

	In this paper, we study deflection of light in static, spherically symmetric, and asymptotically flat spacetimes
	with a Maxwell field coupled with a dilaton field.
	Electrically and/or magnetically charged black hole solutions were found in Ref.\cite{GibbonsMaeda1988,GarfinkleHorowitzea1991} and 
	are sometimes called Garfinkle-Horowitz-Strominger (GHS) solution.
	The electrically charged solution depends on the dilaton coupling constant $\alpha_0\in\mathbb{R}$ (see Sec.\ref{Settings} for the notation)
	as well as the electric charge $Q$ and the mass $M$\footnote{For simplicity, we assume the asymptotic value of the dilaton field
	at spatial infinity is $0$.}.
	GHS solutions include as special cases Gibbons-Maeda-Garfinkle-Horowitz-Strominger (GMGHS) solution $(\alpha_0=1)$,
	Reissner-Nordstr\"{o}m solution $(\alpha_0=0)$, and the solution obtained by Kalza-Klein dimensional reduction
	$(\alpha_0=\sqrt{3})$\cite{GibbonsWiltshire1986,FrolovZelnikovea1987,HorneHorowitz1992}.
	These specific cases as well as the case of $\alpha_0=1/\sqrt{3}$, have been known and studied in the context of string and supergravity theories
	(see, for example, Ref.\cite{Ortin2004} and references therein).

	Deflection of light in a GHS spacetime has been investigated by several authors.
	In Ref.\cite{Bhadra2003}, the gravitational lensing by a GMGHS black hole was studied, and both the weak deflection limit (WDL) and SDL expansions of the deflection angle were derived,
	although the WDL expansion is given as a power series in terms of $M/r_0$ instead of $M/b$,
	where $r_0$ is the radial coordinate at the closest approach of the trajectory.
	For the SDL, the coefficients $\bar{a}$ and $\bar{b}$ were analytically obtained.
	The WDL expansion in terms of $M/b$ for a GMGHS black hole was developed in Ref.\cite{KeetonPetters2005} up to $O((M/b)^3)$.
	For arbitrary values of $\alpha_0$, i.e., GHS black holes, the WDL expansion of the deflection angle was derived in Ref.\cite{OvgunGyulchevea2019} up to $O((M/b)^2)$.
	The SDL for GHS black holes\cite{GyulchevYazadjiev} and its rotating generalization\cite{GyulchevYazadjiev2007} have been reported, although only integral representations are known for $\bar{b}$.

	In Refs.\cite{SasakiSuzuki2021,Sasaki2024}, it was shown that, in the Reissner-Nordstr\"{o}m black hole spacetime, 
	the deflection angle as a function of dimensionless variables $y:=Q^2/4M^2$ and $z:=4M^2/b^2$ satisfies a system of linear partial differential equations,
	called inhomogeneous Picard-Fuchs (PF) equations,
	using which, in addition to the SDL coefficients $\bar{a}$ and $\bar{b}$, higher-order expansions of the deflection angle both in the WDL and SDL were derived.
	It was also pointed out that the integrability condition of the system is a special type of Painlev\'{e} VI equation (PVI)
	that allows an algebraic solution.
	Recently, this method was applied to the Reissner-Nordstr\"{o}m naked singularity spacetime with $Q^2/M^2=9/8$, 
	where the divergence of the deflection angle is not logarithmic
	since the photon sphere is marginally unstable due to the degeneracy with the antiphoton sphere\cite{Sasaki2025}.
	
	In this paper, we attempt to extend the discussion in Refs.\cite{SasakiSuzuki2021,Sasaki2024,Sasaki2025} to GHS spacetimes and
	show that, for specific values of the dilaton coupling constant $\alpha_0\in\{1/\sqrt{3},1,\sqrt{3}\}$, almost the same arguments hold.
	More precisely, for each value of $\alpha_0$, the deflection angle as a function of $s$ and $z$, the former of which is a certain 
	algebraic function of the electric charge to mass ratio $q=Q/M$, is found to obey a system of linear partial differential equations.
	The integrability condition of the system is shown to be a special type of PVIs that allows an algebraic solution.
	We not only describe these mathematical aspects but also show this kind of integrability can be utilized to derive physically relevant quantities.
	In particular, the SDL coefficients $\bar{a}$ and $\bar{b}$\footnote{We use, instead of $\bar{a}$ and $\bar{b}$ themselves, 
	$C_{\rm l}=\bar{a}/2$ and $C_{\rm r}=(\bar{b}+\pi-2\log2)/2$ 
	for notational simplicity in the main discussion. See the discussion around eq. (\ref{dphi_SDLasympt_general}).} 
	are shown to obey 1st order linear differential equations with respect to $\sigma=z_-/z_+$,
	where $z_\pm$ correspond to the critical impact parameters $b^\pm_{\rm c}$ for circular photon orbits as $z_\pm=4M^2/(b^\pm_{\rm c})^2$.
	Note that, in our setting, the parameter $\sigma$ is a function of $q$ so that these differential equations describe the dependence on the background charge
	of the SDL coefficients $\bar{a}$ and $\bar{b}$.
	Interestingly, by using the Hamilton's equations associated with the PVI, 
	the differential equation for $\bar{a}$ can be integrated without specifying explicit dependence on $\sigma$.
	We also discuss a similar structure for another coefficient $\bar{b}$, although the integration needs a case-by-case treatment for each value of the dilaton coupling.

	The rest of the paper consists of four sections; Sec.\ref{Settings} specifies the spacetime metric we study, and an integral representation
	for the deflection angle is derived. Here, we identify the values of the dilaton coupling we study in the next two sections.
	In Sec.\ref{GeneralFormalism}, we then develop our general formalism to analyze the deflection angle
	using the PF equations and introduce the variables for PVI.
	We here derive an expression for $\bar{a}$ in terms of the PVI variables.
	In Sec.\ref{ExplicitAnalysis}, the formalism is applied to each case of the dilaton coupling and the SDL asymptotic expansion for the deflection angle is derived.
	Lastly, we make a concluding remark in Sec.\ref{Conclusion}.

}}}

\section{Spacetime metric and Integral representation of the deflection angle\label{Settings}}
{{{
	The spacetime metric of GHS spacetime is given by\cite{GibbonsMaeda1988,GarfinkleHorowitzea1991}
	\begin{equation}
		\mathrm{d}s^2=-f\mathrm{d}t^2+\frac{\mathrm{d}r^2}{f}+r^2\left(1-\frac{r_-}{r}\right)^{1-\gamma}\mathrm{d}\Omega^2,\label{metric_general}
	\end{equation}
	where 
	\begin{equation}
		f=\left(1-\frac{r_+}{r}\right)\left(1-\frac{r_-}{r}\right)^\gamma,
	\end{equation}
	and $\mathrm{d}\Omega^2=\mathrm{d}\theta^2+\sin^2\theta\mathrm{d}\phi^2$ represents the line element of the unit sphere.
	$r_\pm$ are given in terms of the mass $M$ and the electric charge $Q$ of the spacetime by
	\begin{align}
		r_+=&M\left(1+\sqrt{1-\frac{2\gamma}{1+\gamma}q^2}\right),\\
		r_-=&\frac{M}{\gamma}\left(1-\sqrt{1-\frac{2\gamma}{1+\gamma}q^2}\right),
	\end{align}
	where $q=Q/M$.
	The parameter $\gamma$ is related to the dilaton coupling constant $\alpha_0$ by 
	\begin{equation}
		\gamma=\frac{1-\alpha_0^2}{1+\alpha_0^2}.
	\end{equation}
	Note that as long as $\alpha_0$ is real, $\gamma$ ranges $-1<\gamma\leq1$.
	This metric is a solution to Einstein's equations with a Maxwell gauge field and a dilaton field $\varphi$,
	derived from the following action:
	\begin{equation}
		S=\int\sqrt{-g}\mathrm{d}^4x\left[R-2(\nabla\varphi)^2-e^{-2\alpha_0\varphi}F^2\right],
	\end{equation}
	where $R$ is the Ricci scalar, and $F$ is the field strength of the Maxwell field.

	For $-1<\gamma<1$, $r=r_-$ is a curvature singularity, as can be seen from, for example, the explicit form of the Ricci scalar
	\begin{equation}
		R=-\frac{\gamma^2-1}{2}\frac{r_+r_-^\gamma}{r^{\gamma+3}}\left(\frac{r}{r_+}-1\right)\left(\frac{r}{r_-}-1\right)^{\gamma-2}.\label{Ricci_scalar}
	\end{equation}
	On the other hand, $r=r_+$ is not singular and (if real and larger than $r_-$) represents the event horizon.
	When the electric charge is sufficiently small $0\leq q^2<2/(\gamma+1)$, $r_+>r_-(>0)$ holds and the metric represents a black hole spacetime.

	Due to spherical symmetry, we can focus on the null geodesics on the equatorial plane $\theta=\pi/2$ without loss of generality.
	Then, each null geodesic has the following conserved quantities:
	\begin{align}
		E:=&-g_{tt}\dot{t}=f\dot{t},\label{E}\\
		L:=&g_{\phi\phi}\dot{\phi}=r^2\left(1-\frac{r_-}{r}\right)^{1-\gamma}\dot{\phi},\label{L}
	\end{align}
	where the overdot means differentiation with respect to an affine parameter of the trajectory.
	In addition to these first integrals, we have a constraint $\mathrm{d}s^2=0$, or equivalently
	\begin{equation}
		0=-f\dot{t}^2+\frac{\dot{r}^2}{f}+r^2\left(1-\frac{r_-}{r}\right)^{1-\gamma}\dot{\phi}^2.\label{ds=0}
	\end{equation}
	By eliminating $\dot{t}$ and $\dot{\phi}$ from Eqs.(\ref{E}), (\ref{L}), and (\ref{ds=0}), we obtain
	\begin{equation}
		\dot{r}^2=E^2-\frac{fL^2}{r^2(1-r_-/r)^{1-\gamma}}.\label{rdot}
	\end{equation}
	Combining Eqs.(\ref{L}) and (\ref{rdot}), we obtain 
	\begin{equation}
		\left(\frac{\mathrm{d}\phi}{\mathrm{d}r}\right)^2=\frac{\dot{\phi}^2}{\dot{r}^2}=\frac{L^2}{r^4(1-r_-/r)^{2-2\gamma}}\frac{1}{E^2-fL^2/r^2(1-r_-/r)^{1-\gamma}},
	\end{equation}
	which determines the spatial shapes of trajectories.
	Introducing the dimensionless quantities $u:=2M/r$ and $z:=4M^2E^2/L^2$, this differential equation reduces to
	\begin{equation}
		\left(\frac{\mathrm{d}\phi}{\mathrm{d}u}\right)^2=\frac{1}{z(1-u/u_-)^{2(1-\gamma)}-u^2(1-u/u_+)(1-u/u_-)}.
	\end{equation}
	Considering trajectories of photons incident from and scattered back to infinity,
	the total deflection angle $\hat{\alpha}$ is given by
	\begin{equation}
		\hat{\alpha}=2\Delta\phi-\pi,
	\end{equation}
	where $\Delta\phi$ corresponds to the change in the azimuthal angle $\phi$ between $r=\infty$ and the point of the closest approach $r=r_0$, 
	and has the following integral representation:
	\begin{equation}
		\Delta\phi=\int_0^{u_0}\frac{\mathrm{d}u}{\sqrt{z(1-u/u_-)^{2(1-\gamma)}-u^2(1-u/u_+)(1-u/u_-)}}.\label{integralrep}
	\end{equation}
	The end point of the integration is given by $u_0:=2M/r_0$.
	Note that the parameter $z$ is related to the impact parameter $b$ of the trajectory as $z=4M^2/b^2$.
	In the following, we refer to both $\hat{\alpha}$ and $\Delta\phi$ as the deflection angle.

	In this paper, we restrict our consideration to the specific values of the dilaton coupling $\alpha_0\in\{0,1/\sqrt{3},1,\sqrt{3}\}$,
	where Eq.(\ref{integralrep}) becomes elliptic integrals so that the same method used in Refs.\cite{SasakiSuzuki2021,Sasaki2024} can be applied.
	We note that the spacetimes with these values of $\alpha_0$ are known to be realized as solutions to 
	supergravity theories with a compact space wrapped with various types of branes (see, for example, Ref. cite{Ortin2004} and references therein).
	In particular, when $\alpha_0=0$ the spacetime reduces to the Reissner-Nordstr\"{o}m one.
	Note that although the spacetime metric is the same for the Einstein-Maxwell-Dilaton case and for the supergravity settings,
	gauge field configurations are totally different for these cases.

	Before evaluating the integral (\ref{integralrep}), we have to clarify the physically relevant region in the parameter space.
	Analytic properties of $\Delta\phi$ with respect to $z$ for given background parameters $\alpha_0$ and $q$ are qualitatively different
	depending on the existence of a circular photon orbit.
	The radius $r_{\rm c}$ and the corresponding impact parameter $b_{\rm c}$ of such an orbit are determined by requiring $\dot{r}=\ddot{r}=0$.
	Using the null geodesic equations, we find that $u_{\rm c}:=2M/r_{\rm c}$ solves the following quadratic equation:
	\begin{equation}
		2(\gamma+1)u^2_{\rm c}-\left\{(2\gamma+1)u_++3u_-\right\}u_{\rm c}+2u_+u_-=0.\label{circular_orbits_eq}
	\end{equation}
	We denote the two solutions of this equation by $u_{\rm c}^\pm$.
	The corresponding impact parameters $z_{\pm}:=4M^2/(b^\pm_{\rm c})^2$ are given by
	\begin{equation}
		z_{\pm}=\frac{(u^\pm_{\rm c})^2(1-u^\pm_{\rm c}/u_+)(1-u^\pm_{\rm c}/u_-)}{(1-u^\pm_{\rm c}/u_-)^{2(1-\gamma)}}
		=(u^\pm_{\rm c})^2(1-u^\pm_{\rm c}/u_+)(1-u^\pm_{\rm c}/u_-)^{2\gamma-1}.\label{circular_orbits_z}
	\end{equation}
	Since the spacetime metric (\ref{metric_general}) is asymptotically flat,
	if the smallest real root, say $u^+_{\rm c}$, of Eq.(\ref{circular_orbits_eq}) lies in the interval
	$0<u^+_{\rm c}<\min(u_+,u_-)\ (\Leftrightarrow r^+_{\rm c}>\max(r_+,r_-))$,
	$r=r^+_{\rm c}$ must be unstable circular photon orbits, i.e., a photon sphere\cite{KudoAsada2022}.
	When this is the case, which we assume in the rest of the paper, the region of $z$ where photons can escape to infinity is given by
	$0<z<z_+$.

}}}

\section{General formalism\label{GeneralFormalism}}
{{{
	In this section, we describe our formalism to analytically evaluate the deflection angle given by the integral representation (\ref{integralrep}).
	We restrict our attention to the cases $\alpha_0\in\{0,1/\sqrt{3},1,\sqrt{3}\}$, where the integral (\ref{integralrep}) reduces to elliptic integrals.
	As explicitly shown in Sec.\ref{ExplicitAnalysis} for each value of $\alpha_0$, 
	$\Delta\phi$ as a function of $s$ and $z$, the former of which is a certain algebraic function of the electric charge to mass ratio $q=Q/M$
	,
	obeys a system of linear partial differential equations, called inhomogeneous PF equations,
	\begin{align}
		\left(c_2\partial_z^2+c_1\partial_z+c_0\right)\Delta\phi&=-\dfrac{\rho_0}{\sqrt{z}},\label{PFz}\\
		\left(b_s\partial_s+b_z\partial_z+b_0\right)\Delta\phi&=\tilde{\rho}_0\sqrt{z},\label{PFs}
	\end{align}
	where the coefficients $c_i, b_i, \rho_0$, and $\tilde{\rho}_0$ are polynomials in $s$ and $z$.
	For the derivation of this system, see Ref.\cite{Sasaki2024}.
	In the following, we describe generic features of the PF equations which are common in all the cases of $\alpha_0\in\{0,1/\sqrt{3},1,\sqrt{3}\}$.

	\subsection{Singularities on the $z$-plane}
	Let us consider the first equation (\ref{PFz}).
	Although the explicit expressions of the coefficients $c_i$ are different for each case,
	the differential operator in Eq.(\ref{PFz}) has 5 regular singularities on the $z$-plane for a fixed value of $s$, which we denote
	\begin{equation}
		\left\{0,z_\pm,\Lambda,\infty\right\}.
	\end{equation}
	Among these singularities, $z=z_\pm$ are identical to the critical impact parameter (\ref{circular_orbits_z}) for circular photon orbits,
	one of which, $z_+$, corresponds to, if it exists, the photon sphere,
	while $z=0$ corresponds to the WDL, i.e., $b\to\infty$.
	The head-on limit $z\to\infty\ (b\to0)$ may be relevant when the spacetime is a naked singularity without a photon sphere, 
	i.e., the so-called strongly naked singularity\cite{VirbhadraEllis2002}.
	For the cases we study in this paper, the remaining singularity $z=\Lambda$ turns out to be an apparent one, i.e., any solutions are holomorphic there.

	The differential operator can be cast into the following standard form:
	\begin{equation}
		c_2\partial_z^2+c_1\partial_z+c_0\propto \partial_x^2+p_1\partial_x+p_0=:\mathcal{L},\label{PFz_operator}
	\end{equation}
	where $x:=z/z_+$ is the normalized variable and $p_{1,0}$ are rational functions of $x$ given by
	\begin{equation}
		\begin{cases}
			p_1&=\displaystyle\frac{1}{x}+\frac{1}{x-1}+\frac{1}{x-\sigma}-\frac{1}{x-\lambda},\\
			p_0&=\displaystyle\frac{\kappa}{x(x-1)}+\frac{\lambda(\lambda-1)\mu}{x(x-1)(x-\lambda)}-\frac{\sigma(\sigma-1)H}{x(x-1)(x-\sigma)}.
		\end{cases}
		\label{p1p0}
	\end{equation}
	Here, $\sigma:=z_-/z_+$ and $\lambda:=\Lambda/z_+$ are singularities of the normalized variable, $\kappa$ is a constant, 
	and $\mu,H$ are functions of $s$.
	These functions of $s$ obey the following constraints in order that $z=\Lambda$ is an apparent singularity:
	\begin{equation}
		\sigma(\sigma-1)H=(\lambda-\sigma)\kappa+\lambda(\lambda-1)(\lambda-\sigma)\mu^2+\lambda(\lambda-1)\mu.\label{defH}
	\end{equation}

	From Eq.(\ref{p1p0}), one can read off the characteristic exponents at each singularity, which can be summarized in the following Riemann scheme:
	\begin{equation}
		\left(
		\begin{matrix}
			x=0 & 1 & \sigma & \lambda & \infty \\
			z=0 & z_+ & z_- & \Lambda & \infty\\
			0 & 0 & 0 & 0 & (1+\kappa_\infty)/2 \\
			0 & 0 & 0 & 2 & (1-\kappa_\infty)/2
		\end{matrix}
		\right),\label{Riemann_general}
	\end{equation}
	where $\kappa_\infty^2=1-4\kappa$.
	This, in particular, implies that there exists a unique holomorphic and homogeneous solution normalized to unity at each of $z=0,z_\pm$, 
	which we denote $\omega_{\rm r}^{(0,\pm)}$, respectively.
	Since the characteristic exponents are degenerate there,
	any homogeneous solutions that are independent of $\omega_{\rm r}^{(0,\pm)}$ are logarithmically divergent.
	In particular, this logarithmic divergence at $z=z_+$ corresponds to the log-formula in the SDL\cite{Bozza2002}.

	We note that the Riemann scheme (\ref{Riemann_general}) is valid only when all the singularities are distinctive;
	when any of them coalesce with each other, the characteristic exponents will change.
	Such a coalescence occurs, for example, in the limit $q^2\to9/8$ for Reissner-Nordstr\"{o}m spacetime($\alpha_0=0$).
	In this limit, three of the singularities $z_\pm,\Lambda$ merge to form a single singularity at $z=8/27$, of which the characteristic exponents are $\pm1/6$.
	Physically, this limiting procedure corresponds to the coalescence of a photon and antiphoton spheres 
	to form a marginally unstable photon sphere, and the deflection angle diverges non-logarithmically\cite{Sasaki2025,Tsukamoto2020}.
	Similar phenomena happen for the cases of $\alpha_0$ other than $\alpha_0=0$, which we leave for future work.

	\subsection{Local solutions around $z=0$}
	Since the spacetimes we are considering are asymptotically flat, 
	the total deflection angle $\hat{\alpha}$ vanishes in the WDL.
	Correspondingly, $\Delta\phi$ satisfies the following boundary condition:
	\begin{equation}
		\lim_{z\to+0}\Delta\phi=\frac{\pi}{2}.\label{BC}
	\end{equation}
	Due to the uniqueness of the normalized, holomorphic homogeneous solution $\omega_{\rm r}^{(0)}$ around $z=0$, 
	the deflection angle can be written as
	\begin{equation}
		\Delta\phi=\frac{\pi}{2}\omega_{\rm r}^{(0)}+\Delta\phi_{\rm I}^{(0)}.\label{dphi_WDL}
	\end{equation}
	Here $\Delta\phi_{\rm I}^{(0)}$ is a specific inhomogeneous solution with the asymptotic form $\Delta\phi_{\rm I}^{(0)}=O(\sqrt{z})$ in the limit $z\to0$.
	This asymptotic behavior can be understood from the fact that the inhomogeneous term in Eq.(\ref{PFz}) behaves as $O(1/\sqrt{z})$.
	By using Frobenius's method, one can see that $\omega_{\rm r}^{(0)}$ and $\Delta\phi_{\rm I}^{(0)}$ have the following power series expansions:
	\begin{equation}
		\begin{cases}
			\omega_{\rm r}^{(0)}&\displaystyle=\sum_{n=0}^\infty a_n^{(0)}z^n,\ \ a_0^{(0)}=1\\
			\Delta\phi_{\rm I}^{(0)}&\displaystyle=\sqrt{z}\sum_{n=0}^\infty I_n^{(0)}z^n,
		\end{cases}
		\label{powerseries_z=0}
	\end{equation}
	of which the expansion coefficients $a_n^{(0)}$ and $I_n^{(0)}$ are uniquely determined by Eq.(\ref{PFz}).
	In Sec.\ref{ExplicitAnalysis}, we show that these coefficients are rational functions of $s$ and can be explicitly written in terms of the hypergeometric function ${}_2F_1$.

	\subsection{Local solutions around $z=z_+$ and the SDL formula}
	We next consider local solutions in the SDL $z\to z_+$.
	The Riemann scheme (\ref{Riemann_general}) implies that one can take, as a fundamental set of homogeneous solutions to the first equation (\ref{PFz}),
	the following solutions:
	\begin{equation}
		\begin{cases}
			\omega_{\rm r}^{(+)}=&\displaystyle\sum_{n=0}^\infty a_n^{(+)}\left(1-\frac{z}{z_+}\right)^n,\ \ a_0^{(+)}=1\\
			\omega_{\rm l}^{(+)}=&\displaystyle-\omega_{\rm r}^{(+)}\log\left(1-\frac{z}{z_+}\right)+\sum_{n=1}^\infty b_n^{(+)}\left(1-\frac{z}{z_+}\right)^n.
		\end{cases}
		\label{powerseries_z=z_+}
	\end{equation}
	Here, we assume the principal branch for the $\log$ function.
	Observing that the inhomogeneous term of the first equation (\ref{PFz}) is holomorphic at $z=z_+$, 
	one can see that there exists the following specific solution:
	\begin{equation}
		\Delta\phi_{\rm I}^{(+)}=\sum_{n=1}^\infty I_n^{(+)}\left(1-\frac{z}{z_+}\right)^n.
	\end{equation}
	Again, the expansion coefficients $a_n^{(+)}, b_n^{(+)}$, and $I_n^{(+)}$ are uniquely determined by Eq.(\ref{PFz}).

	Then, general solutions around $z=z_+$ can be written as
	\begin{equation}
		\Delta\phi=C_{\rm r}\omega_{\rm r}^{(+)}+C_{\rm l}\omega_{\rm l}^{(+)}+\Delta\phi_{\rm I}^{(+)},\label{dphi_SDL_general}
	\end{equation}
	where $C_{\rm r, l}$ are $z$-independent arbitrary constants.
	In particular, the asymptotic form in the SDL $z\to z_+$ becomes
	\begin{equation}
		\Delta\phi=-C_{\rm l}\log\left(1-\frac{z}{z_+}\right)+C_{\rm r}+O((1-z/z_+)\log(1-z/z_+)).\label{dphi_SDLasympt_general}
	\end{equation}
	By recalling the definition $z/z_+=(b^+_{\rm c}/b)^2$ and comparing with the log-formula (\ref{SDL_Bozza}),
	one can identify as $\bar{a}=2C_{\rm l}$ and $\bar{b}=2C_{\rm r}+2\log2-\pi$.
	Finding the coefficients $\bar{a}, \bar{b}$, or equivalently $C_{\rm l}, C_{\rm r}$, 
	so that the analytic continuation of $\Delta\phi$ defined around $z=0$ as Eq.(\ref{dphi_WDL}) through the positive real axis 
	matches Eq.(\ref{dphi_SDL_general}) is the main problem to be discussed below.

	\subsection{Integrability conditions} 
	Up to here, we have considered only the first equation (\ref{PFz}).
	Before investigating the effects of the second equation (\ref{PFs}), we discuss the integrability of these two equations.
	Note that if the coefficients of the first equation $\{c_2,c_1,c_0,\rho_0\}$ and those of the second equation $\{b_s,b_z,b_0,\tilde{\rho}_0\}$
	were independently given, then there would not exist any solution.
	Thus, certain integrability conditions must hold among these coefficients.
	Of course, in the cases we study, the existence of a solution is guaranteed because we construct these equations so that a well-defined function (\ref{integralrep}) is a solution to them.
	In spite of this, we describe the integrability conditions in this section
	since they determine the dependence on the background charge $q$ of the SDL coefficients $\bar{a},\bar{b}$, as we show in the next section.

	Integrability of the differential operators in Eqs.(\ref{PFz}) and (\ref{PFs}), 
	especially with the specific form (\ref{PFz_operator}) and (\ref{p1p0}), is known to lead to PVI for the position of the apparent singularity $\lambda$ as a function of $\sigma$\cite{IwasakiKimuraea1991}\footnote{The deformation parameter $\sigma$ is usually denoted by $t$ in the literature, which is not used in this paper in order to avoid 
	confusion with the time coordinate.}.
	PVI can be written in a compact form as a Hamiltonian system,
	\begin{equation}
		\partial_\sigma\lambda=\frac{\partial H}{\partial \mu},\ \ \partial_\sigma\mu=-\frac{\partial H}{\partial\lambda}, \label{Hamiltoneq}
	\end{equation}
	where the Hamiltonian $H=H(\sigma,\lambda,\mu)$ is defined by Eq.(\ref{defH}).
	In our setting, after deriving the coefficients $\{c_2,c_1,c_0\}$ for each case of $\alpha_0\in\{0,1/\sqrt{3},1,\sqrt{3}\}$,
	one can straightforwardly read off a solution $(\lambda,\sigma)$ to a certain type of PVI.
	In particular, $\lambda$ and $\sigma$ are given as rational functions of the parameter $s$, which means that they constitute an algebraic solution to the PVI. 
	In Sec.\ref{ExplicitAnalysis}, we briefly comment on the relations to known algebraic solutions in the literature.

	In addition to PVI, the integrability conditions determine the operator of the second equation (\ref{PFs}) almost uniquely in terms of the variables appearing in Eq.(\ref{p1p0}).
	To explicitly write it down, we change the independent variables as $(s,z)\to(\sigma,x)$. 
	Note that here we use $\sigma=z_-/z_+$ as an independent variable instead of $s$, which is a function of $q=Q/M$.
	Then the operator of Eq.(\ref{PFs}) is transformed to the following standard form:
	\begin{equation}
		b_s\partial_s+b_z\partial_z+b_0\propto \partial_\sigma+q_1\partial_x+q_0=:\mathcal{M},
	\end{equation}
	where $q_{1,0}$ are given by
	\begin{equation}
		\begin{cases}
			q_1&=-\dfrac{(\lambda-\sigma)x(x-1)}{\sigma(\sigma-1)(x-\lambda)},\\
			q_0&=\dfrac{(\lambda-1)(\lambda-\sigma)\mu x}{\sigma(\sigma-1)(x-\lambda)}+\xi(\sigma),
		\end{cases}
	\end{equation}
	with an arbitrary function $\xi(\sigma)$.
	Note that $\xi(\sigma)$ corresponds to the degree of freedom of multiplying an arbitrary function of $\sigma$ to $\Delta\phi$.
	However, in our case, due to the boundary condition (\ref{BC}) that is independent of $\sigma$, this degree of freedom doesn't remain.
	More precisely, by demanding that the operator $\mathcal{M}$ allows $\omega_{\rm r}^{(0)}=1+O(x)$ to be a homogeneous solution around $x=0$, we obtain $\xi(\sigma)=0$.

	\subsection{SDL coefficients via Painlev\'{e} VI Hamiltonian system}
	Applying the operator $\mathcal{M}$ given above to the asymptotic form (\ref{dphi_SDLasympt_general}) 
	and extracting the leading and next-leading order terms in the SDL $x\to1$, we obtain
	\begin{equation}
		\begin{cases}
			\sigma(\sigma-1)(\lambda-1)\partial_\sigma C_{\rm l}-(\lambda-1)(\lambda-\sigma)\mu C_{\rm l}=0,\\
			\sigma(\sigma-1)(\lambda-1)\partial_\sigma C_{\rm r}-(\lambda-1)(\lambda-\sigma)\mu C_{\rm r}=(\lambda-\sigma)C_{\rm l}-\eta(\sigma),
		\end{cases}
		\label{CrCleq}
	\end{equation}
	where 
	\begin{equation}
		\eta(\sigma):=\frac{1}{2\sqrt{z_+}}\left[-\sigma(\sigma-1)\lambda\partial_\sigma z_++(\lambda-\sigma)z_+\right].
	\end{equation}
	Integrating these differential equations, $C_{\rm r,l}$ are fixed up to integration constants, say ${\bf C}_{\rm r,l}$.

	Interestingly, the equation for $C_{\rm l}$ can be integrated without specifying explicit expressions for $\lambda,\mu$, and $H$ as functions of $\sigma$,
	while the integration for $C_{\rm r}$ is carried out in Sec.\ref{ExplicitAnalysis}.
	We use the first equation in Eq.(\ref{Hamiltoneq}), which, with the help of Eq.(\ref{defH}), is explicitly given by
	\begin{equation}
		\partial_\sigma\lambda=\frac{\lambda(\lambda-1)}{\sigma(\sigma-1)}\big\{2(\lambda-\sigma)\mu+1\big\}.
	\end{equation}
	By using this equation, the differential equation for $C_{\rm l}$ can be integrated as
	\begin{equation}
		C_{\rm l}^2={\bf C}_{\rm l}^2\left|\frac{\sigma(\lambda-1)}{\lambda(\sigma-1)}\right|.\label{Cl_general}
	\end{equation}

	The remaining constants ${\bf C}_{\rm r,l}$ are determined by requiring, in the zero-charge limit $q\to0$, that the asymptotic form (\ref{dphi_SDLasympt_general})
	must be consistent with that for the case of a Schwarzschild black hole\cite{BozzaCapozzielloea2001},
	\begin{equation}
		\Delta\phi_{\rm Sch}=-\frac{1}{2}\log\left(\frac{1-z/z_+}{432(2-\sqrt{3})^2}\right)+O((1-z/z_+)\log(1-z/z_+)).\label{SDL_Sch}
	\end{equation}
	As we explicitly see in Sec.\ref{ExplicitAnalysis}, the zero-charge limit corresponds to $|\sigma|\to\infty$ and $|\lambda|\to\infty$.
	Then, this boundary condition leads to ${\bf C}_{\rm l}=1/2$, which results in
	\begin{equation}
		\bar{a}=2C_{\rm l}=\sqrt{\frac{1-1/\lambda}{1-1/\sigma}}. \label{Cl_solution}
	\end{equation}

}}}

\section{Explicit calculations\label{ExplicitAnalysis}}
{{{
	In this section, we apply the formalism developed in the previous section to the cases of $\alpha_0\in\{0,1/\sqrt{3},1,\sqrt{3}\}$ with explicit calculations.
	Each subsection proceeds in the following manner.
	After clarifying the position of the photon sphere $u_{\rm c}^+$, 
	we first show the coefficients of the inhomogeneous PF equations (\ref{PFz}) and (\ref{PFs}),
	from which the positions of singularities $z_\pm,\Lambda$, and the parameters $\sigma,\lambda,\kappa,\mu$, and $H$ are readily obtained.
	At this point, we comment on the relation of the obtained solution $(\lambda,\sigma)$ for PVI to the known solution in the literature.
	We next show the explicit forms of the coefficients $a_n^{(0)}$ and $I_n^{(0)}$ in the power series expansions at $z=0$ in terms of the hypergeometric polynomials.
	The derivation of the PF equations and the proof that $a_n^{(0)}$ and $I_n^{(0)}$ are indeed solutions are not included in this paper
	since these are shown in detail for the case of $\alpha_0=0$, i.e., Reissner-Nordstr\"{o}m spacetime, in Ref.\cite{Sasaki2024}
	and the calculations for other cases are essentially the same.
	Lastly, we consider the connection problem to the SDL $z\to z_+$.
	By employing the generic formula (\ref{Cl_solution}) for $C_{\rm l}$ and solving the second equation of Eq.(\ref{CrCleq}) 
	for $C_{\rm r}$ with the boundary condition given by Eq.(\ref{SDL_Sch}),
	we explicitly derive the SDL formula.
}}}

\subsection{$\alpha_0=0\ (\gamma=1)$\label{alpha=0}}
{{{
	In this case, the metric (\ref{metric_general}) reduces to Reissner-Nordstr\"{o}m one.
	A notable difference from other cases with $\alpha_0\neq0$ is that $r=r_-$ is no longer a curvature singularity, as can be seen from the Ricci scalar (\ref{Ricci_scalar}).
	As a result, the spacetime possesses outer (event) and inner horizons at $r=r_\pm$ when $0<q^2<1$.
	In the extremal limit $q^2\to1$ from below, two horizons coalesce to form a single event horizon.
	If the electric charge exceeds the extremal value $q^2=1$, the event horizon disappears, and the metric represents a naked singularity spacetime.
	
	The quadratic equation for circular photon orbits (\ref{circular_orbits_eq}) has two solutions given by
	\begin{equation}
		u^\pm_{\rm c}=\frac{4}{3(1\pm s)}, \label{uc_RN}
	\end{equation}
	where
	\begin{equation}
		s:=\sqrt{1-\frac{8q^2}{9}}.
	\end{equation}
	One can confirm that $u^-_{\rm c}$ and $u^+_{\rm c}$ are respectively stable and unstable orbits, which form the photon and antiphoton spheres.
	$u_{\rm c}^\pm$ together with the event horizon are plotted in Fig.\ref{RN_PS_fig} as functions of the electric charge $q^2$.
	When the spacetime represents a black hole $(0\leq q^2\leq 1)$, only the photon sphere exists outside the event horizon,
	while the naked singularity is covered by both the photon and antiphoton spheres for $1<q^2<9/8$, i.e., a weakly naked singularity.
	In the limit $q^2\to9/8$ from below, the photon and antiphoton spheres degenerate to form a marginally unstable photon sphere.
	Overcharged cases $q^2>9/8$ then correspond to the strongly naked singularity spacetime.

	The PF equations (\ref{PFz}) and (\ref{PFs}) for the Reissner-Nordstr\"{o}m spacetime have appeared in Ref.\cite{Sasaki2024} for the first time.
	Although the author considered only the black hole case $0<q^2\leq1$ in Ref.\cite{Sasaki2024}, the PF equations are valid even for the naked singularity case.
	Since the notation in this paper is slightly different from that in Ref.\cite{Sasaki2024}, we explicitly show the coefficients below,
	\begin{align}
		&\begin{cases}
			c_2&\displaystyle=512z\bigl[27(s-1)^3z-8(3s-1)\bigr]\bigl[27(s+1)^3z-8(3s+1)\bigr]\\
			&\displaystyle\ \ \ \ \times\bigl[27(s^2-1)(3s^2+1)z-8(9s^2-1)\bigr],\\
			c_1&\displaystyle=1024\bigl[19683(s^2-1)^4(3s^2+1)z^3-2916(s^2-1)(45s^6-15s^4+39s^2-5)z^2\\
			&\displaystyle\ \ \ \ +3456(9s^2-1)(3s^4+6s^2-1)z-256(9s^2-1)^2\bigr],\\
			c_0&\displaystyle=96\bigl[19683(s^2-1)^4(3s^2+1)z^2\\
			&\displaystyle\ \ \ \ -432(s^2-1)(9s^2-1)(27s^4-72s^2+29)z+64(9s^2-1)^2(9s^2+31)\bigr],\\
			\rho_0&\displaystyle=-256\bigl[2187(s^2-1)^3(3s^2+1)z^2\\
			&\displaystyle\ \ \ \ -216(s^2-1)(9s^2-1)(39s^2-7)z-256(9s^2-1)^2\bigr],
		\end{cases}\label{c_alpha0}\\
		&\begin{cases}
			b_s&\displaystyle=27(s^2-1)(3s^2+1)z-8(9s^2-1),\\
			b_z&\displaystyle=-18sz\bigl[9(s^2+3)z-8\bigr],\\
			b_0&\displaystyle=-108sz,\\
			\tilde{\rho}_0&\displaystyle=72s.
		\end{cases}
	\end{align}

	Singularities on the $z$-plane except for $z=0$ and $\infty$ can be read off from $c_2$ as 
	\begin{equation}
		z_\pm=\frac{8(3s\pm1)}{27(s\pm1)^3}, \Lambda=\frac{8(9s^2-1)}{27(s^2-1)(3s^2+1)}.
	\end{equation}
	Note, as mentioned in Sec.\ref{GeneralFormalism}, that $z_\pm$ correspond to the critical impact parameters 
	associated with the circular photon orbits (\ref{uc_RN}) through the formula (\ref{circular_orbits_z}).
	In particular, $z_+$ corresponds to the outer, unstable circular orbits, namely the photon sphere.
	In this paper, we focus on the case of $0<s\leq 1\ (\Leftrightarrow 0\leq q^2<9/8)$, 
	where the deflection angle diverges logarithmically in the SDL $z\to z_+$.
	The boundary case $s=0\ (\Leftrightarrow q^2=9/8)$ was studied in Ref.\cite{Sasaki2025}, 
	where the spacetime possesses a marginally unstable photon sphere and the divergence of $\Delta\phi$ in the SDL is non-logarithmic.

	\begin{figure}[htb]
		\centering
		\includegraphics[keepaspectratio,width=0.49\columnwidth]{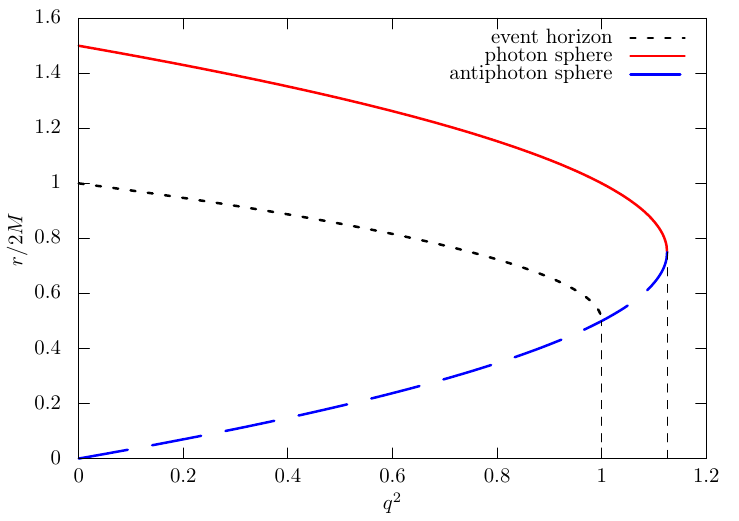}
		\caption{
			The positions of the event horizon $r_+/2M=1/u_+$, the photon sphere $r^+_{\rm c}/2M=1/u_{\rm c}^+$, 
			and the antiphoton sphere $r^-_{\rm c}/2M=1/u_{\rm c}^-$ as functions of the electric charge $q^2$.
			The vertical dashed lines represent boundary cases $q^2=1\ (s=1/3)$ and $q^2=9/8\ (s=0)$.
		}
		\label{RN_PS_fig}
	\end{figure}

	From Eq.(\ref{c_alpha0}), the parameters in Eq.(\ref{p1p0}) are found to be
	\begin{equation}
		\begin{cases}
			\sigma=&\dfrac{(s+1)^3(3s-1)}{(s-1)^3(3s+1)},\\
			\lambda=&\dfrac{(s+1)^2(3s-1)}{(s-1)(3s^2+1)},\\
			\mu=&\dfrac{(1-s)(3s+1)(3s^2+1)}{48s^2(s+1)^2},\\
			H=&\dfrac{(1-s)^3(3s+1)(36s^2+21s+5)}{576s^3(s+1)^3},\\
			\kappa=&\dfrac{3}{16}.
		\end{cases}
		\label{PVIvariables_alpha=0}
	\end{equation}
	As shown in Ref.\cite{Sasaki2024}, $(\lambda,\sigma)$ given above constitutes an algebraic solution to PVI and
	are equivalent, up to certain transformations, to the known solutions in the literature\cite{Hitchin1995,DubrovinMazzocco2000,LisovyyTykhyy2014}.

	The power series expansions with respect to $z$ of $\omega_{\rm r}^{(0)}$ and $\Delta\phi_{\rm I}^{(0)}$ are explicitly given by\cite{Sasaki2024}
	\begin{equation}
		\begin{cases}
			\omega_{\rm r}^{(0)}=&\displaystyle\sum_{n=0}^\infty\frac{(\frac{1}{6})_n(\frac{5}{6})_n}{(n!)^2}\left(\frac{27z}{4}\right)^n{}_2F_1\left[\begin{matrix}-n,-n+\frac{1}{2}\\-3n+\frac{1}{2}\end{matrix};\frac{9(1-s^2)}{8}\right],\\
			\Delta\phi_{\rm I}^{(0)}=&\displaystyle\sqrt{z}\sum_{n=0}^\infty\frac{(\frac{2}{3})_n(\frac{4}{3})_n}{(\frac{3}{2})_n(\frac{3}{2})_n}\left(\frac{27z}{4}\right)^n{}_2F_1\left[\begin{matrix}-n,-n-\frac{1}{2}\\-3n-1\end{matrix};\frac{9(1-s^2)}{8}\right],
		\end{cases}
	\end{equation}
	where $(a)_n:=\Gamma(a+n)/\Gamma(a)$ is the Pochhammer symbol.
	These expressions, together with Eq.(\ref{dphi_WDL}), give a full-order expression for the deflection angle in the WDL $z\to+0$.
	Note that this result in the WDL is valid not only for the black hole case but even for the naked singularity case.

	Next we consider the connection problem to the SDL $z\to z_+$.
	General solutions to the PF equations around $z=z_+$ can be written as Eq.(\ref{dphi_SDL_general}).
	One of the connection coefficients $C_{\rm l}$, or equivalently $\bar{a}$, is given by Eq.(\ref{Cl_solution}), which, in this case, reduces to
	\begin{equation}
		\bar{a}=2C_{\rm l}=\sqrt{\frac{s+1}{2s}}. 
	\end{equation}
	Solving the second equation in Eq.(\ref{CrCleq}), the $s$-dependence of another connection coefficient $C_{\rm r}$ turns out to be
	\begin{equation}
		C_{\rm r}=\sqrt{\frac{s+1}{2s}}\left[{\bf C}_{\rm r}+\frac{1}{2}\log\left(\frac{s^3}{(3s+1)^2}\right)-\log\left(\sqrt{3s+1}+\sqrt{3s}\right)\right].
	\end{equation}
	The remaining constant ${\bf C}_{\rm r}$ is uniquely determined by requiring that the asymptotic formula (\ref{dphi_SDLasympt_general}) reduces,
	in the zero-charge limit $s\to1$, to that for a Schwarzschild black hole (\ref{SDL_Sch}) as
	\begin{equation}
		{\bf C}_{\rm r}=\frac{1}{2}\log6912.
	\end{equation}
	Finally, we obtain an exact expression for the deflection angle $\Delta\phi$ and its asymptotic formula in the SDL $z\to z_+-0$
	\begin{align}
		\Delta\phi=&\sqrt{\frac{s+1}{8s}}\left[\omega_{\rm r}^{(+)}\log\left(\frac{6912s^3}{(3s+1)^2}\left(\sqrt{3s+1}-\sqrt{3s}\right)^2\right)+\omega_{\rm l}^{(+)}\right]+\Delta\phi_{\rm I}^{(+)}\label{SDLexact_alpha=0}\\
		=&-\sqrt{\frac{s+1}{8s}}\log\left(\frac{(3s+1)^2}{6912s^3}\frac{1-z/z_+}{(\sqrt{3s+1}-\sqrt{3s})^2}\right)+O((1-z/z_+)\log(1-z/z_+)).\label{SDLasympt_alpha=0}
	\end{align}
	The SDL formula (\ref{SDLasympt_alpha=0}) was first derived in Ref.\cite{TsukamotoGong2017}, and its higher-order generalization
	(\ref{SDLexact_alpha=0}) was obtained in Ref.\cite{Sasaki2024}.

}}}

\subsection{$\alpha_0=1/\sqrt{3}\ (\gamma=1/2)$\label{alpha=1/sqrt3}}
{{{
	In this case, the positions of the event horizon and the curvature singularity are respectively given by
	$u_+=2M/r_+=2/(1+s)$ and $u_-=2M/r_-=1/(1-s)$, where
	\begin{equation}
		s:=\sqrt{1-\frac{2q^2}{3}}.
	\end{equation}
	The electric charge is restricted to $0\leq q^2\leq 3/2$, otherwise the metric (\ref{metric_general}) becomes imaginary.
	The quadratic equation for circular photon orbits (\ref{circular_orbits_eq}) has two solutions $u_{\rm c}^-=u_-$ and $u_{\rm c}^+=2u_+/3$,
	the latter of which corresponds to the photon sphere when $2u_+/3<u_- \Leftrightarrow 1\geq s>1/7 \Leftrightarrow 0\leq q^2<72/49$.
	Note that the other solution $u_{\rm c}^-$ doesn't give a circular orbit because the areal radius is zero there,
	as can be seen from the metric (\ref{metric_general}).
	\begin{figure}[htb]
		\centering
		\includegraphics[keepaspectratio,width=0.49\columnwidth]{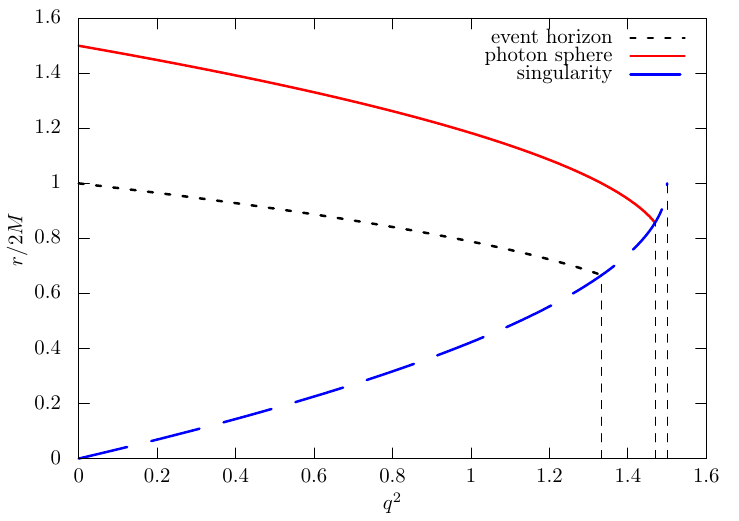}
		\caption{
			The positions of the event horizon $r_+/2M=1/u_+$, the photon sphere $r_{\rm c}^+/2M=1/u_{\rm c}^+$, 
			and the curvature singularity $r_-/2M=1/u_-$ as functions of the electric charge $q^2$.
			The vertical dashed lines correspond to $q^2=4/3\ (s=1/3)$, $q^2=72/49\ (s=1/7)$,
			and $q^2=3/2\ (s=0)$.
		}
		\label{gamma05_horizons}
	\end{figure}
	These positions as functions of the electric charge $q^2$ are plotted in Fig.\ref{gamma05_horizons}.

	The coefficients for the PF equations (\ref{PFz}) and (\ref{PFs}) are found to be
	\begin{align}
		&\begin{cases}
			c_2=&-4z\left[27(s+1)^2z-16\right]\left[2(s-1)^3z-3s+1\right]\left[3(s^2-1)(5s+1)z-4(3s-1)\right], \\
			c_1=&-4\bigl[324(s-1)^4(s+1)^3(5s+1)z^3 \\
				&\hspace{2em}-3(s^2-1)(1213s^4-556s^3+222s^2+692s-275)z^2\\
				&\hspace{2em}+8(3s-1)(113s^3+39s^2+123s-59)z-64(3s-1)^2\bigr],\\
			c_0=&-3\bigl[48(s-1)^4(s+1)^3(5s+1)z^2 \\
				&\hspace{2em}-(s^2-1)(3s-1)(125s^3-157s^2-193s+153)z \\
				&\hspace{2em}+4(3s-1)^2(5s^2+2s+13)\bigr],\\
			\rho_0=&4\bigl[9(s^2-1)^3(5s+1)z^2-18(s^2-1)(s+3)(3s-1)^2z-16(3s-1)^2\bigr], 
		\end{cases}\\
		&\begin{cases}
			b_s=&3(s^2-1)(5s+1)z-4(3s-1),\\
			b_z=&-6z\left[(5s^2+5s+8)z-4\right],\\
			b_0=&-3(s+5)z,\\
			\tilde{\rho}_0=&12.
		\end{cases}
	\end{align}
	From these coefficients, one can read off the singularities on the $z$-plane other than $z=0$ and $z=\infty$ as
	\begin{equation}
		z_+=\frac{16}{27(s+1)^2}, z_-=\frac{1-3s}{2(1-s)^3}, \Lambda=\frac{4(1-3s)}{3(1-s^2)(1+5s)}.
	\end{equation}
	Again, $z_\pm$ correspond to the critical impact parameters given by Eq.(\ref{circular_orbits_z}) with $u_{\rm c}=u_{\rm c}^\pm$, respectively.
	In this paper, we focus on the case $1/7<s\leq 1$,
	where the deflection angle logarithmically diverges in the SDL $z\to z_+$.

	The parameters in the normalized variable defined through Eq.(\ref{p1p0}) are found to be
	\begin{equation}
		\begin{cases}
			\sigma=&\dfrac{27(3s-1)(s+1)^2}{32(s-1)^3},\\
			\lambda=&\dfrac{9(3s-1)(s+1)}{4(s-1)(5s+1)},\\
			\mu=&-\dfrac{2(s-1)(5s+1)}{9(s+1)(7s-1)},\\
			H=&-\dfrac{8(s-1)^3(5s+1)}{9(s+1)^2(7s-1)^2},\\
			\kappa=&\dfrac{2}{9}.
		\end{cases}
	\end{equation}
	$(\lambda,\sigma)$ constitutes an algebraic solution to PVI, which is equivalent to {\it Solution III} in Ref.\cite{LisovyyTykhyy2014}
	\begin{equation}
		\sigma_{\rm III}=\frac{(s_{\rm III}-1)^2(s_{\rm III}+2)}{(s_{\rm III}+1)^2(s_{\rm III}-2)},\lambda_{\rm III}=\frac{(s_{\rm III}-1)(s_{\rm III}+2)}{s_{\rm III}(s_{\rm III}+1)},
	\end{equation}
	with the following transformation:
	\begin{equation}
		\sigma_{\rm III}=1-\frac{1}{\sigma}, \lambda_{\rm III}=1-\frac{\lambda}{\sigma}, s=\frac{2s_{\rm III}-1}{2s_{\rm III}+5}.
	\end{equation}

	The coefficients $a_n^{(0)}$ and $I_n^{(0)}$ for the local solutions (\ref{powerseries_z=0}) at $z=0$ can be explicitly expressed by using the hypergeometric polynomials,
	\begin{equation}
		a_n^{(0)}=\frac{(\frac{1}{6})_n(\frac{5}{6})_n}{(n!)^2}\left(\frac{27(1+s)^2}{16}\right)^n{}_2F_1\left[\begin{matrix}-2n,\frac{1}{2}\\-3n+\frac{1}{2}\end{matrix};\frac{2(1-s)}{1+s}\right],
	\end{equation}
	and
	\begin{equation}
		I_n^{(0)}=\frac{(\frac{2}{3})_n(\frac{4}{3})_n}{(\frac{3}{2})_n(\frac{3}{2})_n}\left(\frac{27(1+s)^2}{16}\right)^n\frac{s+1}{2}{}_2F_1\left[\begin{matrix}-2n-1,\frac{1}{2}\\-3n-1\end{matrix};\frac{2(1-s)}{1+s}\right].
	\end{equation}
	Note that both $\omega_{\rm r}^{(0)}$ and $\Delta\phi_{\rm I}^{(0)}$ solve the 2nd equation (\ref{PFs}) and its homogeneous version as well.

	We next consider the connection problem to the SDL $z\to z_+$.
	One of the connection coefficients $C_{\rm l}$, or equivalently $\bar{a}$, is given by Eq.(\ref{Cl_solution}), which reduces to
	\begin{equation}
		\bar{a}=2C_{\rm l}=\sqrt{\frac{3(s+1)}{7s-1}}.
	\end{equation}
	Solving the second equation in Eq.(\ref{CrCleq}) then gives 
	\begin{equation}
		C_{\rm r}=\sqrt{\frac{3(s+1)}{7s-1}}\left[{\bf C}_{\rm r}+\frac{1}{2}\log\left(\frac{7s-1}{s+5}\right)^2+\log\left\{\frac{7s-1}{s+5}\left(3\sqrt{(s+1)/(7s-1)}-1\right)^2\right\}\right]. \label{Cl_general_gamma=05}
	\end{equation}
	The remaining constant ${\bf C}_{\rm r}$ is fixed by requiring that Eq.(\ref{dphi_SDLasympt_general}) in the zero-charge limit $s\to1$ is consistent with
	Eq.(\ref{SDL_Sch}), as
	\begin{equation}
		{\bf C}_{\rm r}=\frac{1}{2}\log108. \label{Cl_const_gamma=05}
	\end{equation}
	As a result, we obtain the following exact expression and its asymptotic formula in the SDL $z\to z_+-0$:
	\begin{align}
		\Delta\phi=&\frac{1}{2}\sqrt{\frac{3(s+1)}{7s-1}}\left[\omega_{\rm r}^{(+)}\log\left(\frac{1728}{(3\sqrt{(s+1)/(7s-1)}+1)^4}\right)+\omega_{\rm l}^{(+)}\right]+\Delta\phi_{\rm I}^{(+)}\\
		=&-\frac{1}{2}\sqrt{\frac{3(s+1)}{7s-1}}\log\left[\frac{(3\sqrt{(s+1)/(7s-1)}+1)^4}{1728}\left(1-\frac{z}{z_+}\right)\right]+O((1-z/z_+)\log(1-z/z_+)).
	\end{align}

}}}

\subsection{$\alpha_0=1\ (\gamma=0)$\label{alpha=1}}
{{{
	In this case, the positions of the event horizon and the curvature singularity are respectively given by
	$u_+=2M/r_+=1$ and $u_-=2M/r_-=1/s$, where
	\begin{equation}
		s:=\frac{q^2}{2}.
	\end{equation}
	The roots of the quadratic equation for circular photon orbits (\ref{circular_orbits_eq}) are
	\begin{equation}
		u_{\rm c}^{\pm}=\frac{4}{s+3\pm\sqrt{(9-s)(1-s)}}.
	\end{equation}
	One can confirm that $u_{\rm c}^-$ and $u_{\rm c}^+$ are respectively stable and unstable orbits, 
	and that the spacetime possesses a photon sphere at $u_{\rm c}^+$ when $0\leq s\leq 1$.
	\begin{figure}[htb]
		\centering
		\includegraphics[keepaspectratio,width=0.49\columnwidth]{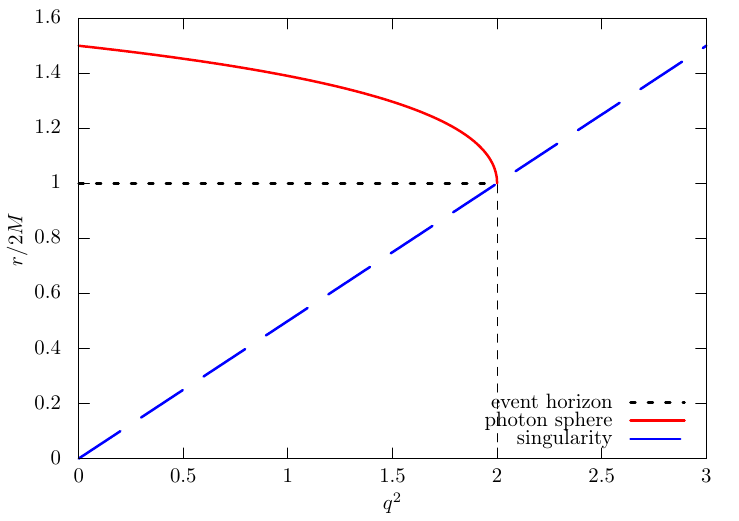}
		\caption{
			The positions of the event horizon $r_+/2M=1/u_+$, the photon sphere $r_{\rm c}^+/2M=1/u_{\rm c}^+$, 
			and the curvature singularity $r_-/2M=1/u_-$ as functions of the electric charge $q^2$.
			The vertical dashed line corresponds to $q^2=2\ (s=1)$.
		}
		\label{gamma0_horizons}
	\end{figure}
	These positions as functions of $q^2$ are plotted in Fig.\ref{gamma0_horizons}.
	In this paper, we focus on the case of $0\leq s<1$.

	The coefficients for the PF equations (\ref{PFz}) and (\ref{PFs}) are found to be
	\begin{align}
		&\begin{cases}
			c_2=&4z(3sz+1)\bigl[4s^3z^2+(s^2+18s-27)z+4\bigr],\\
			c_1=&4\bigl[24s^4z^3+3s(5s^2+18s-27)z^2+2(s^2+18s-27)z+4\bigr],\\
			c_0=&12s^4z^2-3s(3s^2-2s+3)z-(s^2+6s-15),\\
			\rho_0=&-4\bigl[s(s-6)z+1\bigr],
		\end{cases}\\
		&\begin{cases}
			b_s=&-2(s-1)(3sz+1),\\
			b_z=&2z\bigl[(7s-9)z+2\bigr],\\
			b_0=&(s-3)z,\\
			\tilde{\rho}_0=&2,
		\end{cases}
	\end{align}
	from which one can read off the singularities on the $z$-plane other than $z=0,\infty$ as
	\begin{equation}
		z_\pm=\frac{8}{-s^2-18s+27\pm\sqrt{(1-s)(9-s)^3}}, \Lambda=-\frac{1}{3s}.
	\end{equation}
	As in the previous cases, $z_\pm$ correspond to the critical impact parameters given by Eq.(\ref{circular_orbits_z}) with $u_{\rm c}=u_{\rm c}^\pm$, respectively.
	Hereafter, we use the following parameter as well:
	\begin{equation}
		S:=\frac{1}{3}\sqrt{\frac{9-s}{1-s}}.
	\end{equation}
	$z_\pm$ are then written as 
	\begin{equation}
		z_\pm=\frac{(3S\mp1)(3S\pm1)^2}{27(S\pm1)^3}.
	\end{equation}

	The parameters in the normalized variable defined through Eq.(\ref{p1p0}) are found to be
	\begin{equation}
		\begin{cases}
			\sigma=&\dfrac{(S+1)^3(3S-1)}{(S-1)^3(3S+1)},\\
			\lambda=&\dfrac{(S+1)^2}{(1-S)(3S+1)},\\
			\mu=&\dfrac{(S-1)(3S+1)(3S^2+1)}{24S^2(S+1)^2},\\
			H=&\dfrac{(1-S)^3(3S+1)}{576S^3(S+1)^3(3S-1)}(27S^4+108S^3+18S^2+12S-5),\\
			\kappa=&\dfrac{1}{4}.
		\end{cases}
	\end{equation}
	This algebraic solution $(\lambda,\sigma)$ to PVI is one of the so-called Picard solutions\cite{Mazzocco2001}
	\begin{equation}
		\sigma=\frac{(s_{\rm P}-1)^3(s_{\rm P}+3)}{(s_{\rm P}+1)^3(s_{\rm P}-3)}, 
		\lambda=\frac{(s_{\rm P}-1)^2}{(s_{\rm P}-3)(s_{\rm P}+1)},
	\end{equation}
	with the following reparametrization:
	\begin{equation}
		s_{\rm P}=-\frac{1}{s}.
	\end{equation}

	The coefficients $a_n^{(0)}$ and $I_n^{(0)}$ in the power series expansions (\ref{powerseries_z=0}) at $z=0$ can be explicitly expressed as
	\begin{equation}
		a_n^{(0)}=\frac{(\frac{1}{6})_n(\frac{5}{6})_n}{(n!)^2}\left(\frac{27}{4}\right)^n{}_2F_1\left[\begin{matrix}-2n,-n+\frac{1}{2}\\-3n+\frac{1}{2}\end{matrix};s\right],
	\end{equation}
	and
	\begin{equation}
		I_n^{(0)}=\frac{(\frac{2}{3})_n(\frac{4}{3})_n}{(\frac{3}{2})_n(\frac{3}{2})_n}\left(\frac{27}{4}\right)^n{}_2F_1\left[\begin{matrix}-2n-1,-n\\ -3n-1\end{matrix};s\right].
	\end{equation}
	Note that both $\omega_{\rm r}^{(0)}$ and $\Delta\phi_{\rm I}^{(0)}$ solve Eq.(\ref{PFs}) and its homogeneous version as well.

	One of the connection coefficients $C_{\rm l}$, or equivalently $\bar{a}$, is given by Eq.(\ref{Cl_general}), which reduces to
	\begin{equation}
		\bar{a}=2C_{\rm l}=\sqrt{\frac{(S+1)(3S-1)}{4S}}.
	\end{equation}
	Another coefficient can be obtained by solving the second equation in Eq.(\ref{CrCleq}) as
	\begin{equation}
		C_{\rm r}=\sqrt{\frac{(S+1)(3S-1)}{4S}}\left[{\bf C}_{\rm r}+\frac{1}{2}\log\left(\frac{S^3}{(3S-1)^4}\right)+\log\left(\frac{\sqrt{3S}-1}{\sqrt{3S}+1}\right)\right].
	\end{equation}
	The remaining constant ${\bf C}_{\rm r}$ is fixed by requiring that the asymptotic form (\ref{dphi_SDLasympt_general}) in the zero-charge limit $S\to1$
	is equivalent to Eq.(\ref{SDL_Sch}) as
	\begin{equation}
		{\bf C}_{\rm r}=\frac{1}{2}\log6912. 
	\end{equation}
	As a result, we obtain the following exact expression and its asymptotic formula in the SDL $z\to z_+-0$:
	\begin{align}
		\Delta\phi=&\frac{1}{4}\sqrt{\frac{(3S-1)(S+1)}{S}}\left[\omega_{\rm r}^{(+)}\log\left(\frac{6912S^3}{(3S-1)^4}\left(\frac{\sqrt{3S}-1}{\sqrt{3S}+1}\right)^2\right)+\omega_{\rm l}^{(+)}\right]+\Delta\phi_{\rm I}^{(+)}\\
		=&-\frac{1}{4}\sqrt{\frac{(3S-1)(S+1)}{S}}\log\left[\frac{(3S-1)^4}{6912S^3}\left(\frac{\sqrt{3S}+1}{\sqrt{3S}-1}\right)^2\left(1-\frac{z}{z_+}\right)\right]
		+O((1-z/z_+)\log(1-z/z_+)). \label{SDLasympt_alpha=1}
	\end{align}
	We confirmed that the asymptotic formula (\ref{SDLasympt_alpha=1}) is consistent with the SDL formula derived in Ref.\cite{Bhadra2003}.

}}}

\subsection{$\alpha_0=\sqrt{3}\ (\gamma=-1/2)$\label{alpha=sqrt3}}
{{{
	In this case, the positions of the event horizon and the curvature singularity are respectively given by $u_+=2/(1+s)$ and $u_-=1/(s-1)$, where
	\begin{equation}
		s:=\sqrt{1+2q^2}.
	\end{equation}
	The roots of the quadratic equation for circular photon orbits (\ref{circular_orbits_eq}) are 
	\begin{equation}
		u_{\rm c}^\pm=\frac{8}{3(s+1)}\left(1\pm\sqrt{\frac{25-7s}{9(s+1)}}\right)^{-1}.
	\end{equation}
	$u_{\rm c}^+$ corresponds to the photon sphere when $0\leq q^2<4\ (\Leftrightarrow 1\leq s<3)$.
	These positions are plotted in Fig.\ref{gamma-05_horizons} as functions of $q^2$.
	In the following, we focus on the case of $1\leq s<3$.

	\begin{figure}[htb]
		\centering
		\includegraphics[keepaspectratio,width=0.49\columnwidth]{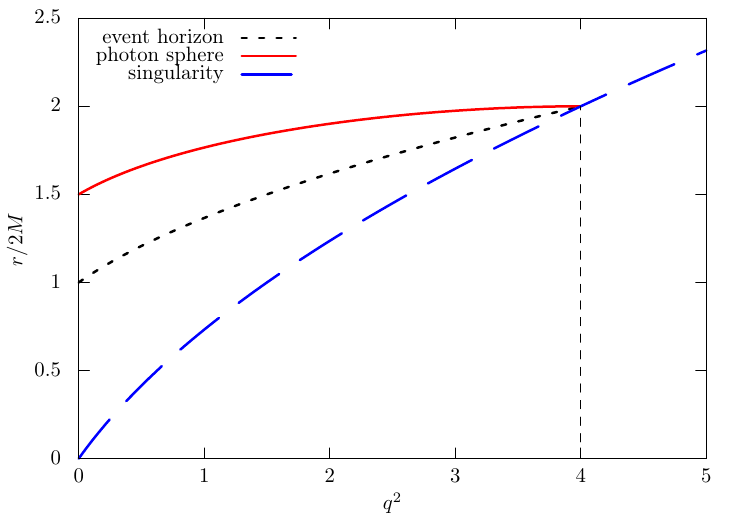}
		\caption{
			The positions of the event horizon $r_+/2M=1/u_+$, the photon sphere $r_{\rm c}^+/2M=1/u_{\rm c}^+$, 
			and the curvature singularity $r_-/2M=1/u_-$ as functions of the electric charge $q^2$.
			The vertical dashed line corresponds to $q^2=4\ (s=3)$.
		}
		\label{gamma-05_horizons}
	\end{figure}

	The coefficients for the PF equations (\ref{PFz}) and (\ref{PFs}) are found to be
	\begin{align}
		&\begin{cases}
			c_2=&4z\left[(s-1)(7-s)z+4\right]\left[-8(3-s)(s-1)^3z^2+(13s^2+10s-131)z+16\right],\\
			c_1=&-4\bigl[16(7-s)(3-s)(s-1)^4z^3-(s-1)(83s^3-399s^2+873s-1205)z^2\\
			    &-8(13s^2+10s-131)z-64\bigr],\\
			c_0=&3(3-s)\left[-2(7-s)(s-1)^4z^2-(s-1)(19s^2-38s+7)z-4(3s+7)\right],\\
			\rho_0=&4\left[(7-s)(3-s)(s-1)^3z^2-2(s-1)(9s^2-28s-5)z-16\right],
		\end{cases}\\
		&\begin{cases}
			b_s=&(3-s)\left[(7-s)(s-1)z+4\right],\\
			b_z=&-2z\left[(s^2-15s+32)z-4\right],\\
			b_0=&-3(3-s)z,\\
			\tilde{\rho}_0=&4.
		\end{cases}
		\label{PFcoeff_KK}
	\end{align}
	Here, we introduece the following parameter:
	\begin{equation}
		S:=\sqrt{\frac{25-7s}{9(s+1)}}.
	\end{equation}
	Then, the singularities on the $z$-plane other than $z=0,\infty$ can be written as
	\begin{equation}
		z_\pm=\frac{(9S^2+7)^2}{108(S\pm1)^3(3S\mp1)},\ \Lambda=\frac{(9S^2+7)^2}{108(S^2-1)(3S^2+1)},
	\end{equation}
	among which $z_\pm$ correspond to the critical impact parameters given by Eq.(\ref{circular_orbits_z}) with $u_{\rm c}=u_{\rm c}^\pm$.

	The parameters in the normalized variable defined through Eq.(\ref{p1p0}) are found to be
	\begin{equation}
		\begin{cases}
			\sigma=&\dfrac{(S+1)^3(3S-1)}{(S-1)^3(3S+1)},\\
			\lambda=&\dfrac{(S+1)^2(3S-1)}{(S-1)(3S^2+1)},\\
			\mu=&\dfrac{(1-S)(3S+1)(3S^2+1)}{48S^2(S+1)^2},\\
			H=&\dfrac{(1-S)^3(3S+1)(36S^2+21S+5)}{576S^3(S+1)^3},\\
			\kappa=&\dfrac{3}{16},
		\end{cases}
	\end{equation}
	which are equivalent to those for the case of $\alpha_0=0$ (see Eq.(\ref{PVIvariables_alpha=0})) under the transformation $s\to S$.

	The coefficients $a_n^{(0)}$ and $I_n^{(0)}$ in the power series expansions (\ref{powerseries_z=0}) can be explicitly expressed as
	\begin{equation}
		a_n^{(0)}=\frac{(\frac{1}{6})_n(\frac{5}{6})_n}{(n!)^2}\left(\frac{27(3-s)(s+1)}{16}\right)^n{}_2F_1\left[\begin{matrix}-n,-n+\frac{1}{2}\\-3n+\frac{1}{2}\end{matrix};\frac{2(s-1)}{s+1}\right],
	\end{equation}
	and
	\begin{equation}
		I_n^{(0)}=\frac{(\frac{2}{3})_n(\frac{4}{3})_n}{(\frac{3}{2})_n(\frac{3}{2})_n}\left(\frac{27(s+1)^2}{16}\right)^n\frac{s+1}{2}{}_2F_1\left[\begin{matrix}-2n-1,-2n-\frac{1}{2}\\-3n-1\end{matrix};\frac{2(s-1)}{s+1}\right].
	\end{equation}

	One of the connection coefficients $C_{\rm l}$, or equivalently $\bar{a}$, is given by Eq.(\ref{Cl_general}), which reduces to
	\begin{equation}
		\bar{a}=2C_{\rm l}=\sqrt{\frac{S+1}{2S}}.
	\end{equation}
	Another coefficient can be obtained by solving the second equation in Eq.(\ref{CrCleq}) as
	\begin{equation}
		C_{\rm r}=\sqrt{\frac{S+1}{2S}}\left[{\bf C}_{\rm r}+\frac{1}{2}\log\left(\frac{S^3}{(3S+1)^2}\right)-\log\left(\frac{\sqrt{6S}+\sqrt{3S-1}}{\sqrt{6S}-\sqrt{3S-1}}\right)\right]. \label{Cl_general_gamma=-05}
	\end{equation}
	The remaining constant ${\bf C}_{\rm r}$ is determined by requiring that asymptotic form (\ref{dphi_SDLasympt_general}) in the zero-charge limit 
	$S\to1$ is equivalent to Eq.(\ref{SDL_Sch}) as
	\begin{equation}
		{\bf C}_{\rm r}=\frac{1}{2}\log6912. \label{Cl_const_gamma=-05}
	\end{equation}
	As a result, we obtain the following exact expression and its asymptotic formula in the SDL $z\to z_+-0$:
	\begin{align}
		\Delta\phi=&\frac{1}{2}\sqrt{\frac{S+1}{2S}}\left[\omega_{\rm r}^{(+)}\log\left(\frac{6912S^3}{(\sqrt{6S}+\sqrt{3S-1})^4}\right)+\omega_{\rm l}^{(+)}\right]+\Delta\phi_{\rm I}^{(+)}\\
		=&-\frac{1}{2}\sqrt{\frac{S+1}{2S}}\log\left[\frac{(\sqrt{6S}+\sqrt{3S-1})^4}{6912S^3}\left(1-\frac{z}{z_+}\right)\right]+O((1-z/z_+)\log(1-z/z_+)).
	\end{align}

}}}

\section{Conclusion\label{Conclusion}}
{{{
	In this paper, we studied the deflection angle of light in GHS spacetimes with the specific values of the dilaton coupling constant $\alpha_0\in\{0,1/\sqrt{3},1,\sqrt{3}\}$.
	We showed that the deflection angle satisfies a system of inhomogeneous Picard-Fuchs equations with respect to $z=4M^2/b^2$ and $s$,
	of which the integrability condition corresponds to Painlev\'{e} VI equation.
	By using Hamilton's equation associated with it, we obtained an expression of one of the SDL coefficients $\bar{a}$,
	i.e., Eq.(\ref{Cl_solution}), in terms of the critical impact parameters corresponding to circular photon orbits through $\sigma=(b^+_{\rm c}/b^-_{\rm c})^2$,
	as well as the apparent singularity $\lambda$ of the PF equation.
	We also derived a differential equation for the other SDL coefficient $\bar{b}$ with respect to the background electric charge,
	which, together with the SDL formula for a Schwarzschild black hole, determines the exact expression of $\bar{b}$.
	In particular, such expressions for the cases with $\alpha_0=1/\sqrt{3}$ (Eq.(\ref{Cl_general_gamma=05}) with Eq.(\ref{Cl_const_gamma=05})) 
	and $\alpha_0=\sqrt{3}$ (Eq.(\ref{Cl_general_gamma=-05}) with Eq.(\ref{Cl_const_gamma=-05}))
	appear for the first time in the literature,
	while those for Reissner-Norstr\"{o}m $(\alpha_0=0)$ and GMGHS $(\alpha_0=1)$ cases recover the known results\cite{TsukamotoGong2017,Bhadra2003}.
	We also succeeded in deriving full order expressions of the deflection angle in the WDL, i.e.,  $a_n^{(0)}$ and $I_n^{(0)}$.

	Compared with the standard method for calculating the SDL coefficients $\bar{a}$ and $\bar{b}$ developed by Refs.\cite{Bozza2002,Tsukamoto2017},
	our method has several advantages.
	First, one can derive higher-order expansions in systematic manners both for the WDL and SDL 
	because the expansion coefficients obey linear recurrence relations due to the existence of linear differential equations. 
	Second, once the PF equations, which are linear differential equations, are derived, 
	subsequent analysis has less ambiguity than the standard method.
	For example, in the standard method, separation of the exact integral into a divergent part $I_{\rm D}$ and a remainder $I_{\rm R}$ is not unique,
	while there is no necessity for such separation in deriving the PF equations.

	On the other hand, our calculation relies heavily on the PF equations,
	which exist only when the deflection angle is written as an integral on a smooth family of algebraic varieties.
	Elliptic curves are the simplest cases, where the PF equations become 2nd order\footnote{
		The order of the PF equations could be higher than two when the integrand has poles other than the branch points.
	}.
	If the algebraic curve associated with a deflection angle has genus larger than 1, 
	the order of the PF equations becomes higher than 2.
	In the case of GHS spacetimes with $\alpha_0\in\mathbb{R}$, if $2(1-\gamma)$ is a rational number,
	the deflection angle given by Eq.(\ref{integralrep}) can be cast into a hyperelliptic integral by changing the integration variable.
	Then, the PF equations would be of order $2g$, where $g$ is the genus of the hyperelliptic curve.
	Although explicit calculations would be cumbersome for such cases, similar analysis with the case of elliptic integrals is possible at least in principle.
	However, if the dilaton coupling value $\alpha_0$ takes a value such that $2(1-\gamma)$ is an irrational number,
	our method is no longer applicable.
}}}

\printbibliography
\end{document}